\documentclass[12pt]{article}

\usepackage{amssymb}
\usepackage{amsmath}
\usepackage{amscd}
\usepackage{latexsym}
\usepackage{graphicx}

\usepackage{cite}

\newcommand{\be}{\begin{equation}}
\newcommand{\ee}{\end{equation}}

\newcommand{\dlt}{\delta}
\newcommand{\prt}{\partial}
\newcommand{\br}{{\bf r}}

\newcommand{\vp}{\varphi}

\newcommand{\al}{\alpha}
\newcommand{\ra}{\rightarrow}
\newcommand{\sgm}{\sigma}

\newcommand{\Om}{\Omega}

\newcommand{\dgr}{\dagger}

\newcommand{\rgl}{\rangle}
\newcommand{\lgl}{\langle}

\begin{document}

\begin{center}

{\Large{\bf Optical lattice with heterogeneous atomic density} \\ [5mm]

V.I. Yukalov$^{1}$ and E.P. Yukalova$^{2}$ } \\ [3mm]

{\it
$^1$Bogolubov Laboratory of Theoretical Physics, \\
Joint Institute for Nuclear Research, Dubna 141980, Russia \\ [3mm]

$^2$Laboratory of Information Technologies, \\
Joint Institute for Nuclear Research, Dubna 141980, Russia }

\vskip 2mm

E-mail: yukalov@theor.jinr.ru

\end{center}

\vskip 0.5cm

\begin{abstract}

The possibility is considered for the formation in optical lattices of a
heterogeneous state characterized by a spontaneous mesoscopic separation 
of the system into the spatial regions with different atomic densities. 
It is shown that such states can arise, if there are repulsive interactions 
between atoms in different lattice sites and the filling factor is less than 
one-half.

\end{abstract}

\vskip 5mm


\newpage

\section{Introduction}

Optical lattices, loaded with cold atoms, are intensively studied, being the 
objects with rich properties that can be widely regulated (see, e.g., the 
review articles \cite{Morsh_1,Moseley_2,Bloch_3,Yukalov_4}). The loaded atoms 
can interact with each other through short-range as well as long-range forces,
such as dipolar forces \cite{Griesmaier_5,Baranov_6,Baranov_7}. 

In addition to usual insulating and delocalized equilibrium states, atoms in 
optical lattices can form several quasi-equilibrium and metastable states. 
For example, in optical lattices, there can exist metastable states with 
repulsively bound bosonic pairs \cite{Winkler_8}, metastable states 
characterized by microscopic phase separation in a mixture of two bosonic 
species \cite{Roscilde_9}, quasi-equilibrium mixture of localized and 
itinerant bosons \cite{Yukalov_10}, and metastable states of atoms with 
dipolar interactions \cite{Santos_11}. Double-well optical lattices can 
display the states with mesoscopic disorder, characterized by a heterophase 
mixture of mesoscopic regions with ordered and disordered atomic imbalance 
\cite{Yukalov_12,Yukalov_13}. Incorporating into the system of cold atoms 
impurities \cite{Massignan_14} or imposing random external fields 
\cite{Sanchez_15} can produce glassy lattice states \cite{Gurarie_16} 
similar to vitrified solid states of metals \cite{Johnson_17}. 

In the present paper, we consider the possibility of forming in an optical 
lattice of a heterophase state consisting of regions with different atomic 
densities. These regions have mesoscopic spatial sizes and are randomly 
distributed in space, where they are not fixed, but can appear and disappear
in different places. In that sense, such a state is a dynamical heterophase 
mixture analogous to other heterophase states with mesoscopic phase separation,
which occur in many condensed-matter systems \cite{Yukalov_18,Yukalov_19}.
Each subregion of a competing phase is a kind of a droplet, or grain, of 
a denser phase inside a diluted phase. Such states are, of course, not 
absolutely equilibrium, but are quasi-equilibrium.  

The typical linear size of a dense droplet is defined by the length
$l_{cor}$, at which atoms are strongly correlated and can coherently form 
a single phase. This length is mesoscopic, being between the mean interatomic 
distance $a$ and the linear system size $L$,
$$
 a \ll l_{cor} \ll L \;  .
$$
The droplet size is rather of nanoscale, not exceeding the critical radius,
after which the germ would grow, provoking a phase transition in the whole 
system \cite{Bakai_20}. Nanoscale nuclei of a competing phase are not 
equilibrium and, strictly speaking, thermodynamic notions, such as surface
tension or surface energy, may be not applicable \cite{Kharlamov_21}. The 
lifetime $t_{cor}$ of a correlated subregion, forming a droplet, is also 
mesoscopic, being between the local equilibration time $t_{loc}$ and the 
observation time $t_{obs}$,
$$
 t_{loc} \ll t_{cor} \ll t_{obs} \;   .
$$
Generally speaking, the sizes and lifetimes of the droplets are of multiscale
nature, being inside mesoscopic intervals, for which $l_{cor}$ and $t_{cor}$
play the role of centers \cite{Yukalov_22}. To some extent, the denser 
subregions remind the grains arising in the process of grain turbulence
\cite{Yukalov_23}. An opposite situation happens in the case of a solid with 
cracks and pores, where there are low-density regions inside a more dense solid 
\cite{Yukalov_24}.

A snapshot of the heterophase two-density state is shown in Fig. 1, where 
the regions of higher density are randomly located inside a matrix of lower 
density. 

The consideration of a new thermodynamic state necessarily includes the 
analysis of its stability. Analyzing this, we show that there exist 
conditions, when the two-density state in an optical lattice is really 
stable. These conditions, briefly speaking, require the presence of 
intersite atomic interactions and a low filling factor, smaller than 
one-half.

\section{Heterophase two-density lattice state}

The first step for treating a two-phase system with random subregions is 
the averaging over heterophase configurations \cite{Yukalov_18,Yukalov_19}.
Keeping in mind the standard form of the Hamiltonian, after averaging over 
configurations, we come to the effective Hamiltonian 
\be
\label{1}
\widetilde H = H_1 \bigoplus H_2 \;   ,
\ee
consisting of two terms
\be
\label{2}
H_\al = w_\al \int \psi_\al^\dgr(\br) \left [ \hat H_L(\br) - 
\mu \right ] \psi_\al(\br) \; d\br \; + \; 
\frac{w_\al^2}{2} \int \psi^\dgr_\al(\br) \psi^\dgr_\al(\br') \Phi(\br-\br')
\psi_\al(\br') \psi_\al(\br) \; d\br d\br'
\ee
representing two different phases, whose atoms are described by the field
operators $\psi_\alpha$, with $\alpha = 1, 2$. Here $\hat{H}_L(\bf r)$ is
an optical-lattice Hamiltonian and $\Phi(\bf r)$ is a pair interaction 
potential. The Hamiltonian is renormalized by the geometric phase 
probabilities
\be
\label{3}
 w_\al \equiv \frac{V_\al}{V} \qquad ( V_1 + V_2 = V ) \;  ,
\ee
where $V_\alpha$ is the average volume occupied by the $\alpha$ - phase and 
$V$ is the system volume. By this definition, the phase probability satisfies
the properties
\be
\label{4}
 w_1 + w_2  = 1 \; , \qquad 0 \leq w_\al \leq 1 \;  .
\ee

By assumption, the phases have different densities 
\be
\label{5}
 \rho_\al \equiv \frac{N_\al}{V_\al} = \frac{1}{V}
 \int \lgl \psi^\dgr_\al(\br) \psi_\al(\br) \rgl \; d\br \;  ,
\ee
in which the number of atoms in an $\alpha$ - phase is
\be
\label{6}
N_\al = w_\al \int \lgl \psi^\dgr_\al(\br) \psi_\al(\br) \rgl \; d\br \; .
\ee
Without the loss of generality, we may call the first phase more dense,
so that
\be
\label{7}
 \rho_1 > \rho_2 \;  .
\ee
In that sense, the densities, distinguishing the phases play the role of the
order parameters. 
  
The optical lattice prescribes the spatial periodicity of the lattice 
Hamiltonian $\hat{H}_L(\bf r)$ with respect to the lattice vectors enumerated
by the index $j = 1,2,\ldots,N_L$ running through all $N_L$ lattice sites. 
The field operators can be represented as expansions 
\be
\label{8}
\psi_\al(\br) = \sum_{nj} e_j^\al c_{nj} \vp_{nj}(\br)
\ee
over the localized orbitals $\varphi_{nj}$. The expansion takes into account 
that a $j$ - site can be either occupied by an atom or free, depending on the 
value of the variable $e_j^\alpha = 0,1$.  

We assume that each lattice site can host not more than one atom, which is 
expressed through the unipolarity condition
\be
\label{9}
 \sum_n c^\dgr_{nj} c_{nj} = 1 \; , \qquad 
c_{mj}^\dgr c_{nj}^\dgr = 0 \;  .
\ee

Substituting expansion (8) into Hamiltonian (2) yields two types of terms, 
with respect to the site indices $i$ and $j$. The terms, describing atomic 
interactions, define the effective time $t_{osc}$ of atomic oscillations in
the vicinity of a given site. The other type of the terms is responsible 
for the hopping of atoms between the lattice sites, which can be characterized 
by a hopping time $t_{hop}$. The observation time has to be much longer than 
the hopping time, so that various phase configurations could be realized in 
the system, thus, justifying the averaging over these configurations. The 
relation between $t_{osc}$ and $t_{hop}$ describes whether the system is in 
an insulating or delocalized state. When $t_{osc}$ is much shorter than 
$t_{hop}$, the atoms are well localized. This implies that the interaction 
terms are much larger than the hopping terms, responsible for atom hopping. 
In what follows, we assume that atoms are sufficiently well localized, so that 
the hopping terms are small, as compared to the interaction terms. Then the 
diagonal approximation can be employed corresponding to the following form of 
the matrix elements:
\be
\label{10}
 \lgl m i \; | \; \hat H_L \; | \; n j \rgl = 
\dlt_{mn} \dlt_{ij} E_0 \; , \qquad
\lgl m i , n j \; | \; \Phi \; | \; m' i', n' j' \rgl = 
\dlt_{mm'} \dlt_{nn'} \dlt_{ii'} \dlt_{jj'} \Phi_{ij} \;  .
\ee
The constant term $E_0$ can be incorporated into the chemical potential. In 
this way, Hamiltonain (2) reduces to the form
\be
\label{11}
H_\al =  - w_\al \mu \sum_{j=1}^{N_L} e_j^\al \; + \;
\frac{1}{2} \; w_\al^2 \sum_{i\neq j}^{N_L} \Phi_{ij} e_i^\al e_j^\al \; .
\ee
  
Using the canonical transformation
\be
\label{12}
e_j^\al = \frac{1}{2} + S_j^z \qquad \left ( e_j^\al = 0,1 \right ) \; ,
\qquad  \qquad
S_j^z = e_j^\al - \; \frac{1}{2} \qquad 
\left (  S_j^z  = \pm \frac{1}{2} \right ) \; ,
\ee
we come to the pseudospin representation
\be
\label{13}
H_\al = \frac{N_L}{8} \; \left ( w_\al^2 \Phi - 4 w_\al \mu \right ) + 
\frac{1}{2}\; \left ( w_\al^2 \Phi - 2 w_\al \mu \right ) \sum_{j=1}^{N_L} S_j^z \; + \;
\frac{w_\al^2}{2} \sum_{i\neq j}^{N_L} \Phi_{ij} S_i^z S_j^z \;  ,
\ee
where
\be
\label{14}
 \Phi \equiv \frac{1}{N_L} \sum_{i\neq j}^{N_L} \Phi_{ij} = 
\sum_{j(\neq i)}^{N_L} \Phi_{ij} \;  .
\ee

Then we resort to the mean-field approximation resulting in the Hamiltonian
\be
\label{15}
 H_\al = \frac{N_L}{8} \; \left [ w_\al^2 \Phi \left ( 1 - s_\al^2 \right ) -
4 w_\al \mu \right ] + \frac{1}{2} \; 
\left [ w_\al^2 \Phi \left ( 1 + s_\al \right ) - 2 w_\al \mu \right ]
\sum_{j=1}^{N_L} S_j^z \;  ,
\ee
in which the notation
\be
\label{16}
s_\al \equiv 2 \lgl S_j^z \rgl_\al = 
\frac{2}{N_L} \sum_{j=1}^{N_L} \; \lgl S_j^z\rgl_\al
\ee
is introduced. The average $\langle S_j^z \rangle_\alpha$ is taken with 
respect to Hamiltonain (15). Quantity (16) can be calculated either directly 
or by minimizing the thermodynamic grand potential
\be
\label{17}
 \frac{\Om}{N_L} = \frac{1}{8} \sum_\al w_\al^2 \Phi(1-s_\al^2) - \; 
\frac{1}{2}\; \mu  -  2T\ln 2  -  T \sum_\al \ln \; \cosh \left [
\frac{w_\al^2\Phi(1+s_\al) - 2 w_\al\mu}{4T} \right ] \;  ,
\ee
which gives
\be
\label{18}
s_\al = \tanh  \left [
\frac{ 2 w_\al\mu - w_\al^2\Phi(1+s_\al) }{4T} \right ] \;  .
\ee

Minimizing the grand potential (17) with respect to the phase probability,
under the normalization condition (4), we set
\be
\label{19}
 w \equiv w_1 \; , \qquad w_2 = 1 - w \;  .
\ee
The minimization yields the equation
$$
 \sum_\al \left [ 2\mu s_\al - w_\al \Phi (1 + s_\al)^2 \right ] \;
\frac{\prt w_\al}{\prt w} = 0 \;  ,
$$
from which we may express the chemical potential 
\be
\label{20}
\mu = \frac{w_1(1+s_1)^2 - w_2(1+s_2)^2}{2(s_1-s_2)} \; \Phi \;   .
\ee

An important quantity is the filling factor
\be
\label{21}
 \nu \equiv \frac{N}{N_L} \qquad ( N = N_1 + N_2 ) \;  .
\ee
Representing the particle number (6) as
\be
\label{22}
 N_\al = \frac{N_L}{2} \; w_\al (1 + s_\al) \;  ,
\ee
we have the filling factor
\be
\label{23}
\nu = \frac{1}{2} \sum_\al w_\al ( 1 + s_\al ) \; .
\ee
Inverting this with respect to the phase probability of the dense phase, we 
get
\be
\label{24}
 w_1 = \frac{2\nu -1 -s_2}{s_1 - s_2} \;  .
\ee 

It is convenient to introduce the dimensionless quantity
\be
\label{25}
 x_\al \equiv \frac{1}{N_L} \sum_{j=1}^{N_L} \; \lgl e_j^\al \rgl_\al \; ,
\ee
playing the role of a dimensionless order parameter. For the dense and 
diluted phases, we write
\be
\label{26}
 x \equiv x_1 = \frac{1}{2}\; (1 + s_1) \; , \qquad  
y \equiv x_2 = \frac{1}{2}\; (1 + s_2) \; ,
\ee
respectively. The density of the $\alpha$-phase reads as
\be
\label{27}
\rho_\al \equiv \frac{N_\al}{V_\al} = \frac{N_L}{V} \; x_\al \;   ,
\ee
which shows why quantities (25) play the role of dimensionless order 
parameters. Due to inequality (7), we have the condition
\be
\label{28}
 x > y \;  ,
\ee
distinguishing the phases with respect to their densities.  

The filling factor (23) can be written as
\be
\label{29}
 \nu =  w_1 x + w_2 y \;  .
\ee
Because of condition (28), the relation
\be
\label{30}
y \leq \nu \leq x
\ee
holds true. 

Using notation (26) reduces the chemical potential (20) to the form
\be
\label{31}
 \mu = \frac{w_1 x^2 - w_2 y^2}{x-y} \; \Phi \;  .
\ee
From Eq. (24), we find the phase probabilities
\be
\label{32}
 w_1 = \frac{\nu-y}{x-y} \; , \qquad w_2 = \frac{x-\nu}{x-y} \;  .
\ee

The sign 
\be
\label{33}
\sgm \equiv \frac{\Phi}{|\Phi|} = {\rm sgn} \Phi
\ee
of the effective interaction (14), for a while, is arbitrary. Measuring
temperature in units of $\vert \Phi \vert$, for the order-parameters (26), 
we get
$$
2x =  1 + \sgm \tanh \left [ \frac{w_1y ( w_1x - w_2y) }{2T(x-y) }
\right ] \; ,
$$
\be
\label{34}
 2y =  1 + \sgm \tanh \left [ \frac{w_2x ( w_1x - w_2y) }{2T(x-y) }
\right ] \;  .
\ee

Thermodynamic quantities can be found from the free energy, for which we 
define the dimensionless quantity
\be
\label{35}
 F \equiv \frac{\Om+\mu N}{N_L|\Phi|} = 
\frac{\sgm}{2} \left [ w_1^2 x(1-x) + w_2^2 y (1-y) \right ]  + 
\frac{T}{2} \ln [ x (1-x)y(1-y) ]  + 
\frac{\mu}{|\Phi|} \left ( \nu -\; \frac{1}{2} \right ) \;  .
\ee

\section{Stability of heterophase two-density state}

First of all, we recall that to be stable a heterophase system has to satisfy
the necessary heterophase stability condition
\be
\label{36}
 \left ( \left \lgl \frac{\prt^2 H}{\prt w^2} \right \rgl \right ) > 0 \;  ,
\ee
which follows from the minimization of the grand potential 
\cite{Yukalov_18,Yukalov_19}. This leads to the inequality
\be
\label{37}
  \left \lgl \frac{\prt^2 H}{\prt w^2} \right \rgl = 
N_L \Phi (x^2 + y^2 ) > 0\;  ,
\ee
from which it is clear that the effective interaction (14) has to be 
effectively repulsive, so that 
\be
\label{38}
  \Phi > 0 \; .
\ee
An effectively attractive interaction does not allow for the formation of 
a stable heterophase system. 

Additionally, the system should be thermodynamically stable, implying that 
the specific heat
\be
\label{39}
C_V = - T  \left ( \frac{\prt^2 F}{\prt T^2} \right )_V
\ee
and the isothermal compressibility
\be
\label{40}
\kappa_T = \frac{1}{\nu^2} \;  
\left ( \frac{\prt^2 F}{\prt \nu^2} \right )^{-1}_T
\ee
be non-negative and finite \cite{Yukalov_25}, 
\be
\label{41}
 0\leq C_V < \infty \; , \qquad 0 \leq \kappa_T < \infty \;  .
\ee

In what follows, we again measure temperature in units of $\Phi$. Keeping 
in mind that we need to consider only repulsive atomic interactions, we 
have to solve the system of equations for the order parameter of the dense 
phase
\be
\label{42}
 2x =  1 + \tanh \left \{ \frac{wy [ wx - (1-w)y]}{2T(x-y) } \right \} \;  ,
\ee
the order parameter of the rarefied phase
\be
\label{43}
2y =  1 + \tanh \left \{ \frac{(1-w)x [ wx - (1-w)y]}{2T(x-y) } \right \}\;  ,
\ee
and the probability of the dense phase
\be
\label{44}
 w = \frac{\nu-y}{x-y} \;  .
\ee
The corresponding solutions define the free energy
$$
F = \frac{1}{2} \left [ w^2 x (1-x) + (1-w)^2 y (1-y) \right ] + 
\frac{1}{2}\; T \ln [ x (1-x) y(1-y) ] \; +
$$
\be
\label{45}
 + \; \left ( \nu -\; \frac{1}{2} \right ) \;
\frac{wx^2-(1-w)y^2}{x-y} \;   ,
\ee
from where the specific heat and compressibility can be calculated. 

Solving Eqs. (42), (43), and (44), we are looking for the probability 
in the interval $0 \leq w \leq 1$ and for the order parameters satisfying
the inequalities 
$$
0 \leq y \leq \nu \leq x \leq 1 \; .
$$
Numerical investigation shows that the heterophase system can be stable 
only for small filling factors,
$$
 0 < \nu < \frac{1}{2} \;  .
$$
For larger filling factors, compressibility (40) becomes negative, although 
specific heat (39) is always positive.     

Solutions to Eqs. (42) to (44) exist in the temperature interval 
$[T_n, T_n^*]$. These temperatures can be called the lower nucleation 
temperature and upper nucleation temperature. The lower nucleation 
temperature $T_n$ is found numerically, being zero for $\nu < 0.32$. The 
upper nucleation temperature $T_n^*$ is defined by the conditions
$$
w(T_n^*) = 0 \; , \qquad x(T_n^*) = \frac{1}{2} \; , \qquad
y(T_n^*) = \nu \;   ,
$$
which yields
$$
T_n^* = \frac{\nu}{(1-2\nu)\ln(\frac{1}{\nu}-1)} \;   .
$$
When $\nu \ra 1/2$, then both $T_n$ and $T_n^*$ tend to infinity. Table 1 
gives the values of $T_n$ and $T_n^*$ in the allowed interval of 
$0 < \nu < 1/2$.  

Figure 2 presents the behavior of solutions as functions of temperature for 
different filling factors and Fig. 3, as functions of the filling factor for
different temperatures. Specific heat (39) and compressibility (40) are 
positive. For illustration, $C_V$, as a function of temperature, is shown 
in Fig. 4.

\section{Comparison with pure single-density state}

The heterophase two-density state should be compared with the pure 
single-phase state, when $w \equiv 1$. Then the grand potential is
\be
\label{46}
 \frac{\Om_1}{N_L} = \frac{1}{8} \; \Phi (1 - s^2 ) \; - \; 
\frac{1}{2}\; \mu_1 \; - \;
T\ln 2 \; - \; T\ln \; \cosh \left [ \frac{\Phi(1+s)-2\mu_1}{4T} \right ] \;  ,
\ee
where
\be
\label{47}
 s = \tanh \left [ \frac{2\mu_1-\Phi(1+s)}{4T} \right ] \;  .
\ee
The filling factor reads as
\be
\label{48}
\nu \equiv \frac{N}{N_L} =\frac{1}{2}\; ( 1+s) \;  .
\ee
From Eqs. (47) and (48), we have
$$
2\nu = 1 + \tanh \left ( \frac{\mu_1 -\nu \Phi}{2T} \right ) \;   ,
$$
which results in the chemical potential
\be
\label{49}
\mu_1 = \nu \Phi + T \ln \frac{\nu}{1-\nu} \;   .
\ee

For the dimensionless free energy, we get
\be
\label{50}
 F_1 \equiv \frac{\Om_1+\mu_1N}{N_L|\Phi|} = 
\frac{1}{2}\;\sgm \nu^2 + T [ \nu \ln\nu +
(1-\nu)\ln (1-\nu) ] \;  .
\ee
This shows that, for the pure phase, the specific heat is zero, $C_V = 0$.
And for the compressibility, we find
$$
\kappa_1 = \frac{1-\nu}{\nu[\sgm\nu(1-\nu)+T]} \;   .
$$
The latter is positive, provided that
$$
T > -\sgm \nu (1-\nu) \;   .
$$
In the case of repulsive interactions, when $\sigma = 1$, the pure phase 
can exist at all temperatures. But for attractive interactions, when 
$\sigma = -1$, the system is stable only for sufficiently high temperatures,
such that $T > \nu (1 - \nu)$.  
 
Comparing the free energy (45) of the heterophase two-density state with 
the free energy (50) of the pure single-density state, we find that, in all 
those cases, when the heterophase state exists, $F < F_1$. This is shown
in Fig. 5 for repulsive interactions. Therefore, in the temperature region 
$T_n < T < T_n^*$, the heterophase state is stable, while the pure state 
is metastable.

\section{Conclusion}

We have considered the possibility of the formation in optical lattices of 
a heterogeneous state characterized by a spontaneous mesoscopic separation 
of the system into the spatial regions with two different atomic densities,
one being more dense than the other. We show that such states can really 
occur, provided that atomic interactions between atoms in different lattice 
sites are repulsive and the filling factor is less than one-half. The 
heterophase state is stable in the temperature region between the lower, $T_n$,
and upper, $T_n^*$, nucleation temperatures.

\section*{Acknowledgement}

Financial support from the Russian Foundation for Basic Research (grant
14-02-00723) is appreciated.

\newpage

\begin{center}
{\Large{\bf Figure Captions} }
\end{center}

\vskip 3cm

{\bf Figure 1}. Snapshot of a heterophase two-density lattice system. Regions of
higher density $\rho_1$ are randomly immersed into the matrix of lower
density $\rho_2$, with $\rho_1 > \rho_2$.

\vskip 1cm
{\bf Figure 2}. Solutions as functions of dimensionless temperature $T$ for 
different filling factors: (a) order parameters $x$ (solid line) and $y$ 
(dashed line) for $\nu = 0.1$ (line 1), $\nu = 0.2$ (line 2), $\nu = 0.3$ 
(line 3), and $\nu = 0.32874$ (line 4); (b) dense-phase probability $w$ for 
the same filling factors and enumeration as above; (c) order parameters $x$ 
(solid line) and $y$ (dashed line) for $\nu = 0.33$ (line 1), $\nu = 0.4$ 
(line 2), and $\nu = 0.45$ (line 3); (d) dense-phase probability $w$ for 
the same filling factors and enumeration as in (c). 

\vskip 1cm
{\bf Figure 3}. Order parameters $x$ (solid line) and $y$ (dashed line) and the 
dense-phase probability $w$ (dashed-dotted line) as functions of filling 
factor $\nu$ for different temperatures: (a) $T = 0.01$; (b) $T = 0.5$.  

\vskip 1cm
{\bf Figure 4}. Specific heat as function of temperature for different filling 
factors: (a) $\nu = 0.1$ (line 1), $\nu = 0.2$ (line 2), $\nu = 0.3$ 
(line 3), and $\nu = 0.32874$ (line 4); (b) $\nu = 0.33$ (line 1), $\nu = 0.4$ 
(line 2), and $\nu = 0.45$ (line 3).  

\vskip 1cm
{\bf Figure 5}. Free energy $F$ (solid line) of the heterophase state, compared 
to the free energy $F_1$ (dashed line) of the pure state, for varying 
temperature and different filling factors: (a) $\nu = 0.3$; (b) $\nu = 0.45$.

\newpage

\begin{center}
{\Large {\bf Table Caption}}
\end{center}

\vskip 1cm
{\bf Table 1}. Lower and upper temperatures defining the existence interval of 
the optical lattice with heterogeneous densities.  

\vskip 3cm

\begin{center}
{\bf Table 1}

\vskip 3mm
\begin{tabular}{|c|c|c|} \hline
$\nu$ & $T_n$ & $T^*_n$ \\  \hline
0.1     & 0        & 0.05689 \\ \hline
0.2     & 0        & 0.24045 \\ \hline
0.3     & 0        & 0.88517 \\ \hline
0.32874 & 0.01     & 1.34442 \\ \hline
0.33    & 0.01253  & 1.37053 \\ \hline
0.4     & 0.520515 & 4.93261 \\ \hline
0.45    & 3.9583   & 22.4248 \\ \hline
\end{tabular}

\end{center}

\newpage

%Figure 1
\begin{figure}[ht]
\centerline{\includegraphics[width=12cm]{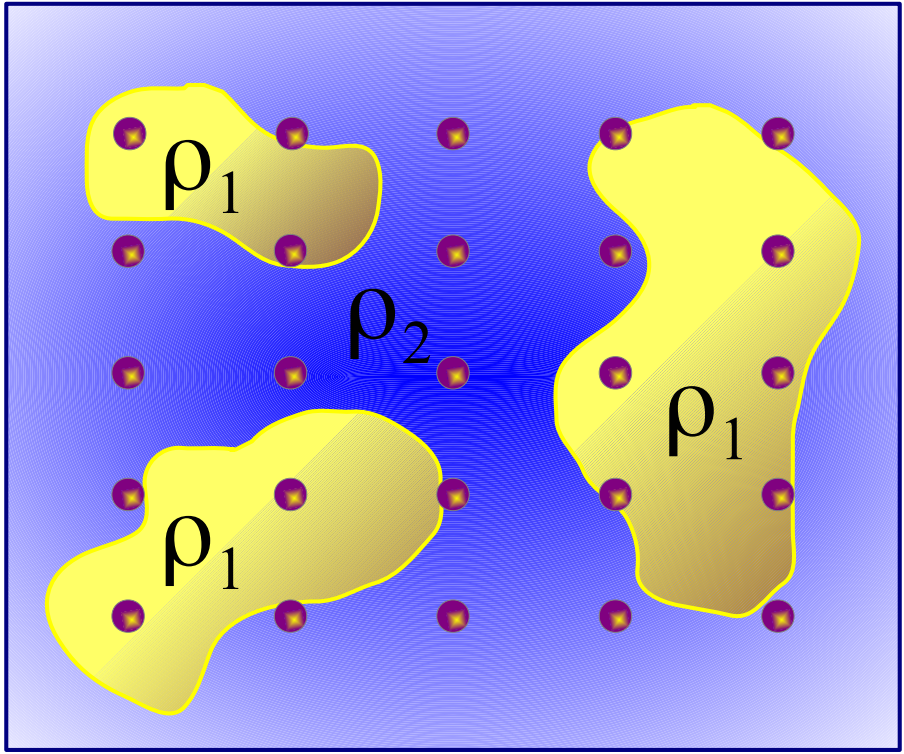} }
\caption{Snapshot of a heterophase two-density lattice system. Regions of
higher density $\rho_1$ are randomly immersed into the matrix of lower
density $\rho_2$, with $\rho_1 > \rho_2$. 
}
\label{fig:Fig.1}
\end{figure}

\newpage

%Figure 2
\begin{figure}[ht]
\vspace{9pt}
\centerline{
\hbox{ \includegraphics[width=8.5cm]{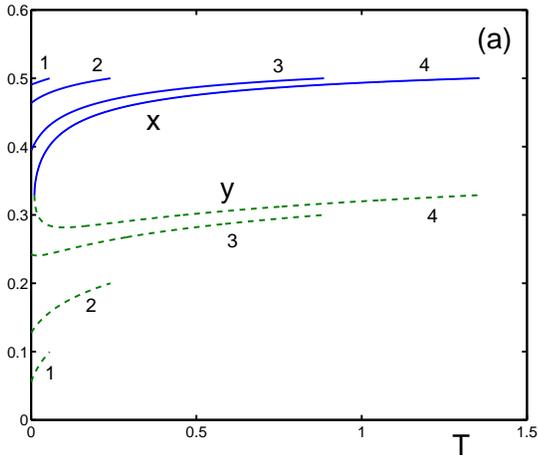} \hspace{1cm}
\includegraphics[width=8.5cm]{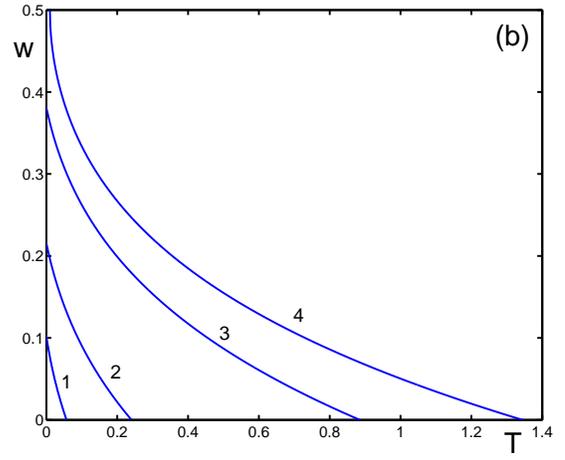}  } }
\vspace{9pt}
\centerline{
\hbox{ \includegraphics[width=8.5cm]{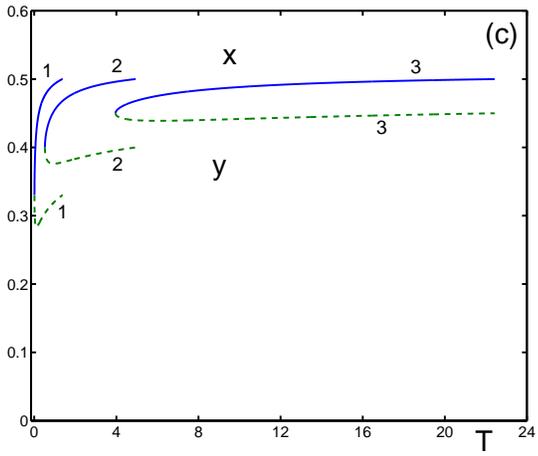} \hspace{1cm}
\includegraphics[width=8.5cm]{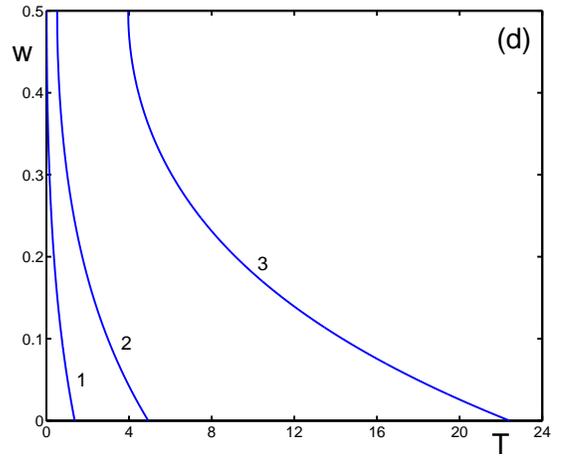} } }
\caption{Solutions as functions of dimensionless temperature $T$ for 
different filling factors: (a) order parameters $x$ (solid line) and $y$ 
(dashed line) for $\nu = 0.1$ (line 1), $\nu = 0.2$ (line 2), $\nu = 0.3$ 
(line 3), and $\nu = 0.32874$ (line 4); (b) dense-phase probability $w$ for 
the same filling factors and enumeration as above; (c) order parameters $x$ 
(solid line) and $y$ (dashed line) for $\nu = 0.33$ (line 1), $\nu = 0.4$ 
(line 2), and $\nu = 0.45$ (line 3); (d) dense-phase probability $w$ for 
the same filling factors and enumeration as in (c). 
}
\label{fig:Fig.2}
\end{figure}

\newpage

%Figure 3
\begin{figure}[ht]
\vspace{9pt}
\centerline{
\hbox{ \includegraphics[width=8.5cm]{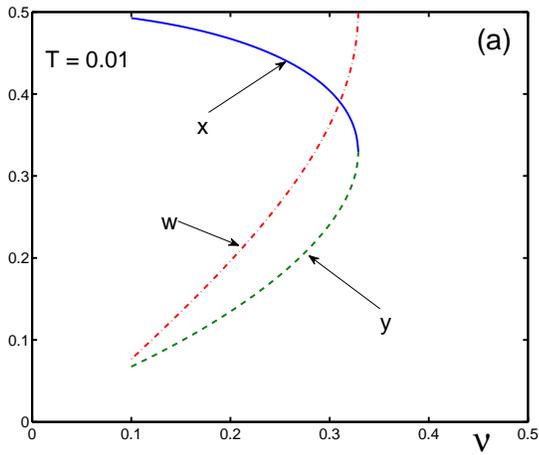} \hspace{1cm}
\includegraphics[width=8.5cm]{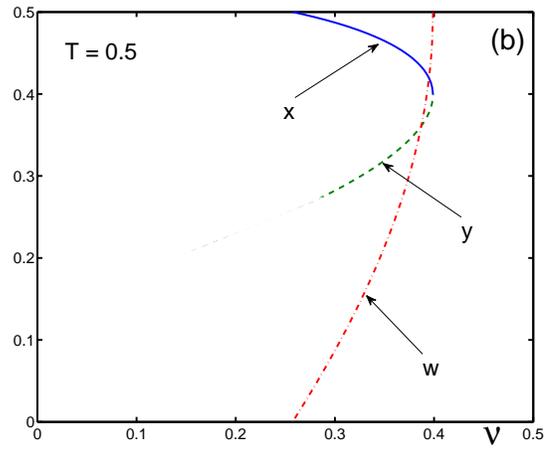} } }
\caption{Order parameters $x$ (solid line) and $y$ (dashed line) and the 
dense-phase probability $w$ (dashed-dotted line) as functions of filling 
factor $\nu$ for different temperatures: (a) $T = 0.01$; (b) $T = 0.5$.  
}
\label{fig:Fig.3}
\end{figure}

\newpage

%Figure 4
\begin{figure}[ht]
\vspace{9pt}
\centerline{
\hbox{ \includegraphics[width=8.5cm]{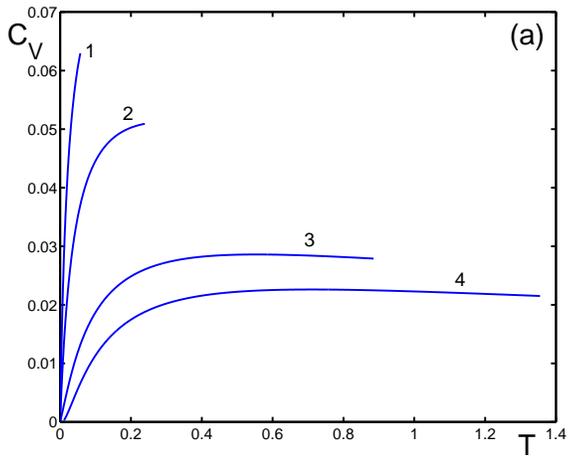} \hspace{1cm}
\includegraphics[width=8.5cm]{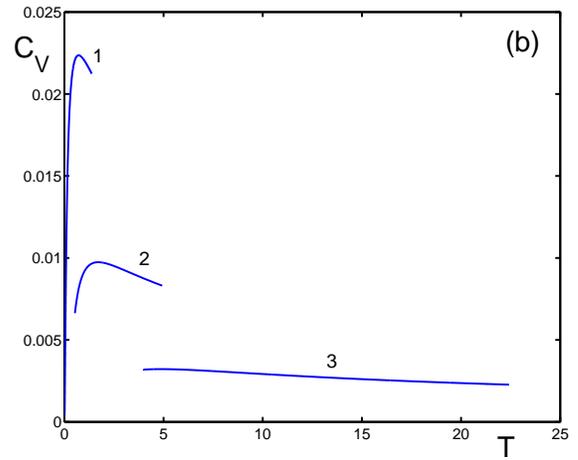} } }
\caption{ Specific heat as function of temperature for different filling 
factors: (a) $\nu = 0.1$ (line 1), $\nu = 0.2$ (line 2), $\nu = 0.3$ 
(line 3), and $\nu = 0.32874$ (line 4); (b) $\nu = 0.33$ (line 1), $\nu = 0.4$ 
(line 2), and $\nu = 0.45$ (line 3).  
}
\label{fig:Fig.4}
\end{figure}

\newpage

%Figure 5
\begin{figure}[ht]
\vspace{9pt}
\centerline{
\hbox{ \includegraphics[width=8.5cm]{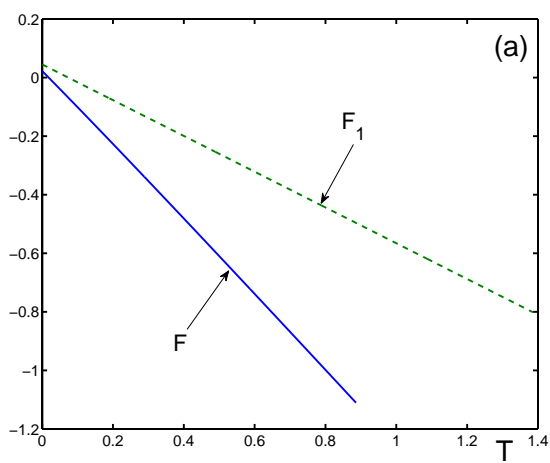} \hspace{1cm}
\includegraphics[width=8.5cm]{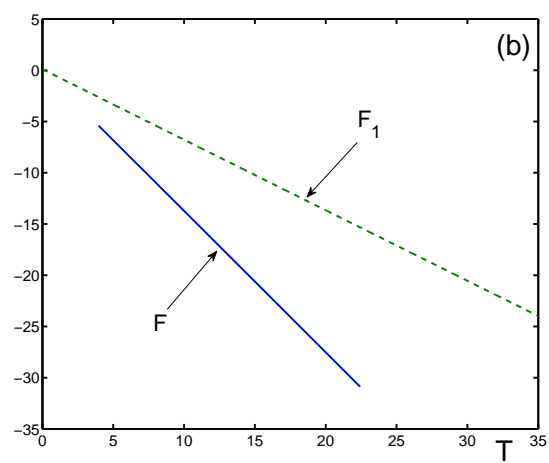} } }
\caption{Free energy $F$ (solid line) of the heterophase state, compared 
to the free energy $F_1$ (dashed line) of the pure state, for varying 
temperature and different filling factors: (a) $\nu = 0.3$; (b) $\nu = 0.45$. 
}
\label{fig:Fig.5}
\end{figure}

\end{document}